# Active Loss Engineering in Vanadium Dioxide Based BIC Metasurfaces

Andreas Aigner[1], Filip Ligmajer[2,3,*], Katarína Rovenská[2,3], Jakub Holobrádek[2], Beáta Idesová[2,3], Stefan A. Maier[1,4,5], Andreas Tittl[1,*], and Leonardo de S. Menezes[1,6]

1) Chair in Hybrid Nanosystems, Nano-Institute Munich, Faculty of Physics, Ludwig-Maximilians-University Munich, Munich, 80539, Germany.
2) Central European Institute of Technology, Brno University of Technology, 61200 Brno, Czech Republic
3) Institute of Physical Engineering, Faculty of Mechanical Engineering, Brno University of Technology, 61669 Brno, Czech Republic.
4) School of Physics and Astronomy, Monash University, Clayton, Victoria 3800, Australia.
5) Department of Physics, Imperial College London, London SW7 2AZ, United Kingdom.
6) Department of Physics, Universidade Federal de Pernambuco, 50670-901 Recife-PE, Brazil

*E-mails: Filip.Ligmajer@vutbr.cz; andreas.tittl@physik.uni-muenchen.de

**ABSTRACT**

Metasurfaces have unlocked significant advancements across photonics, yet their efficient active control remains challenging. The active materials required often lack continuous tunability, exhibit inadequate refractive index (RI) changes, or suffer from high losses. These aspects pose an inherent limitation for resonance-shifting based switching: when RI changes are small, the resulting shift is also minor. Conversely, high RI changes typically come with high intrinsic losses necessitating broad modes because narrow ones cannot tolerate such losses. Therefore, larger spectral shifts are required to effectively detune the modes. This paper introduces a novel active metasurface approach that converts the constraint of high intrinsic losses into a beneficial feature. This is achieved by controlling the losses in a hybrid vanadium dioxide ($VO_2$) – silicon metasurface, supporting symmetry-protected bound states in the continuum (BICs) within the infrared spectrum. By leveraging the temperature-controlled losses in $VO_2$ and combining them with the inherent far-field-coupling tunability of BICs, we gain unprecedented precision in independently controlling both the radiative and nonradiative losses of the resonant system. Our dual-control mechanism allows us to optimize our metasurfaces and we experimentally demonstrate quality factors above 200, a maximum reflectance amplitude of 90%, a relative switching contrast of 78%, and continuous tuning from under- to over-coupling within the infrared spectral range. This study provides a foundation for experimentally and technologically simple, fine-tunable, active metasurfaces for applications ranging from molecular sensors to filters and optical modulators.

## INTRODUCTION

Metasurfaces[1,2] have achieved remarkable advancements in photonics, including flat lenses[3,4], holograms[5,6], molecular sensors[7,8], lasers[9,10], and photocatalytic reactors[11,12]. However, their predominantly passive nature severely limits their practical use. Consequently, the focus has shifted towards developing active solutions[13,14,15] that manipulate the refractive index (RI) of materials through external stimuli like electrical tuning[16,17], optical illumination[18,19], or temperature[20,21]. Apart from metasurfaces designed for quick, high contrast modulation for switches[17] and lenses[22], a new class of tunable metasurfaces is emerging. These prioritize the precise adjustment of amplitude, shape, and spectral position of resonances. Yet, they often fall short due to the challenges associated with the materials used. Materials with minor RI changes often lack the desired tunability[23], while those with substantial RI changes exhibit high intrinsic losses like germanium-antimony-telluride[24,25], indium-antimony-telluride[18,26], conductive polymers[27,28], and vanadium-dioxide ($VO_2$)[29,30]. This limitation complicates the commonly used detuning-based switching method, which relies on shifting a resonance towards larger or smaller wavelengths. Spectrally sharp resonances cannot sustain high losses due to their long resonance lifetime. While broad resonances are capable of handling high losses and optically or electrically switched metasurfaces show great promise[31,32], their spectral width and high loss limits their applicability in many applications demanding high-transmission like optical filtering. Additionally, the need for spectral selectivity in multiplexing applications is not met by these broad resonances.

Recently, symmetry-protected bound states in the continuum (BICs)[33,34] have gained attention in the metasurface community. Their appeal lies in their spectrally narrow resonances quantified by the quality (Q) factor, which is the ratio of resonance frequency $\omega_0$ to resonance bandwidth $\delta\omega$, and in the straightforward tunability of far-field coupling through simple geometric parameters. This led to demonstrations of numerous exciting physical phenomena with applications for lasers[35], chiral resonators[36], molecular sensors[37], and lenses[38]. Applying the principle of BICs to tunable metasurfaces is, however, experimentally challenging. Popular tunable materials like GST, conductive polymers, or $VO_2$ have not yet been achieved experimentally due to the susceptibility of high Q-factor resonances to intrinsic losses[39]. Therefore, active BICs have only been realized in the THz range[40] or through low RI changing materials[19].

Here, we propose a new approach for achieving actively fine-tunable metasurfaces based on loss-engineered hybrid $VO_2$-Si resonators[41,31,42] exhibiting BICs. Rather than fighting the high intrinsic losses in $VO_2$'s high-temperature phase[30,29], we utilize it to generate a loss-based active tuning effect. Because the temperature induced phase change in $VO_2$ starts locally in sub-wavelength areas[43], the overall phase transition is gradual, with intermediate states readily accessible within a temperature window of roughly 30°C. When combined with the BIC's asymmetry-based tunability of radiative losses, this approach provides us with two tuning mechanisms: a passive one via the asymmetry of the structure and an active one via the temperature (Fig. 1a). This dual control allows precise tailoring for various potential applications which prioritize high Q-factors, large resonance amplitudes, strong field enhancements, high modulation, or a combination of these key parameters. Our numerical and experimental demonstrations reveal that for a conventional, ellipse-pair based BIC metasurface geometry, a thin $VO_2$ layer of about 30 nm is sufficient to quench the resonance. The remainder of the 750 nm high resonator can comprise any low-loss dielectric; in our case, it includes 600 nm of silicon below and 120 nm of SiO2 on top of the VO2 (see a sketch of the structure in Fig. 1b and the scanning electron microscope (SEM) image of the fabricated structure in Fig. 1c). With 2 in the low-temperature phase, we achieve Q-factors of up to 200 and a maximum reflectance amplitude of 90%. By heating the metasurface to temperatures above the phase transition, we achieve a relative switching amplitude of up to 78% between the "cold" and the "hot" phase (Fig. 1d,e). We experimentally mapped the parameter space defined by the asymmetry parameter $\alpha$ and temperature $T$, which govern radiative and intrinsic losses, respectively. Using temporal coupled mode theory (TCMT) modelling, we identified the impact of our dual-control mechanism on both loss rates. Additionally, our results confirm the structure's versatility in transitioning between all coupling states, ranging from under-coupling through critical coupling to over-coupling.

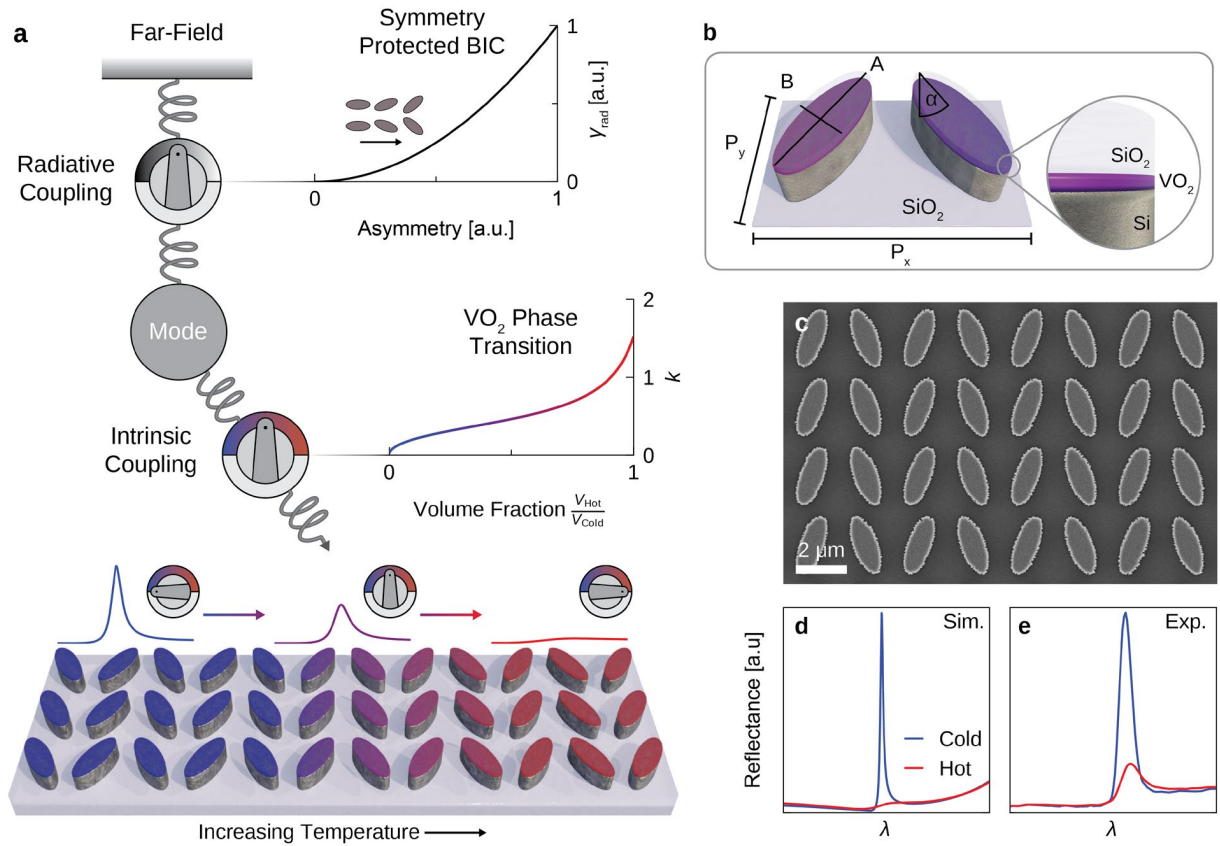

**Figure 1: Tuning principle and resonator geometry. (a)** Schematics of the active VO2-Si BIC metasurface. The resonator is coupled to incoming and outgoing radiation via the asymmetry-dependent radiative loss $\gamma_{\text{rad}}$, as depicted for the tilted ellipse-pair geometry where $\gamma_{\text{rad}} \propto \sin^2(\alpha)$ with the tilting angle

$\alpha$. Additionally, the mode is coupled to intrinsic losses which can be fine-tuned via the gradual phase transition of VO$_2$ and its increasing imaginary part $k$ of the complex RI. This change in $k$ can gradually supress a resonant mode. **(b)** Sketch of the tilted ellipse-pair unit cell defined by the ellipses' long axis $A$, the short axis $B$, the tilting angle $\alpha$, the pitch in x-direction $P_x$ and in y-direction $P_y$. The resonator, on top of a SiO$_2$ substrate, consists of a Si layer, a VO$_2$ layer, and a capping SiO$_2$ layer. **(c)** SEM image of the fabricated metasurface with $\alpha = 20°$. Sketch of **(d)** numerical and **(e)** experimental reflectance spectra for the VO$_2$-Si metasurface in its hot and cold phases.

## RESULTS

We chose a tilted ellipse-pair geometry for our metasurface design due to its remarkable resonance robustness[44]. Furthermore, compared to other designs, this geometry offers a notably low spectral reflectance baseline of 2-3% in simulations. This feature, along with an exceptionally clear off-resonant spectrum, makes it particularly suitable for switching applications that value spectral cleanliness, high switching contrast, and multiplexing capability. While we focus on the tilted ellipse pair, it is worth noting that our methodology is not constrained to this specific BIC metasurface geometry and can be applied to other geometries.

To implement our loss-based switching concept, we need an active material that satisfies two main requirements. In its "off" state, the losses of the material should be minimal to ensure that resonant modes are not damped. Conversely, in its "on" state, the material should have a very high absorbance to achieve optimal resonance quenching. VO$_2$, especially in the mid-IR, stands out as an extraordinary choice, meeting both criteria[29,30]. Our atomic layer deposited VO$_2$ (see Methods section) exhibits at 6.5 μm an extinction coefficient $k$ of 0.011 at 25°C, and a $k$ of 2.42 at 80°C, as validated via ellipsometry (see Fig. S1).

Using simulations (see Methods section for details), we tuned the geometric parameters of our structures to resonate between 6-7 μm. This leads to a 750 nm high resonator on top of a SiO$_2$ substrate, which is composed of a Si layer at the bottom, a VO$_2$ layer in the middle, and a 120 nm SiO$_2$ capping layer which acts as a mask during the VO$_2$-Si etching.

While SiO2 has some absorption in the mid-IR range, it was chosen as substrate material since its thermal expansion coefficient aligns well with Si, ensuring a stable film during the VO2 annealing process. We sourced the RIs for VO2 through ellipsometry, shown in Fig. S1, and took the values for Si and SiO2 from Shkondin et al.[45] and Kischkat et al.[46], respectively. To model the intermediate states of the VO2 layer, where it is partially in the cold and partially in the hot phase, we employed an effective medium approximation[47] (Supplementary Note 1) with a hot-phase volume fraction $VF = \frac{V_{\text{hot}}}{V_{\text{total}}}$, defined as the ratio of volume in the hot phase $V_{\text{hot}}$ divided by the total volume $V_{\text{total}}$.

Resonance detuning-based switching typically requires a resonator predominantly composed of an active medium, because it changes the effective cavity size through refractive index shifts. Conversely, our loss-based approach only needs a minimal portion of the medium to be active. With this method, the effective cavity size hardly changes, instead, we merely introduce intrinsic losses into the system to attenuate the mode. By examining the relationship between the ellipses' tilting angle $\alpha$, the VO2 layer height, and the resulting reflectance modulation, we observed absolute reflectance modulation amplitudes $\max(R_{\text{cold}} - R_{\text{hot}})$ of up to 80% (see Fig. 2a). Note that this corresponds to a maximum relative modulation $\max\left(\frac{R_{\text{cold}} - R_{\text{hot}}}{R_{\text{cold}}}\right)$ of 97% (Fig. S2). Notably, as the asymmetry increases, the reflectance modulation peaks between $\alpha = 15 - 30°$, after which it declines. We ascribe this trend to two counteracting effects: the resonance lifetime and the absorption of VO2 in its cold state. At smaller asymmetries with high Q-factors $Q$, the longer resonance lifetime (given by $\tau = \frac{Q}{\omega_0}$) allows light to be in extended contact with the VO2 layer and thereby the absorption is enhanced. However, VO2 also exhibits absorption when in its cold state ($k = 0.011$ at 6.5 µm), which results in a lower overall reflectance amplitude for small asymmetries in the cold phase. The same effects also lead to the observed behaviour for different VO2 layer heights: thin VO2 layers achieve optimal modulation with small asymmetries, and thicker layers favour larger asymmetries. Due to the dampening influence of a thicker VO2 layer on both the Q-factor and the reflectance amplitude in the cold phase, we proceed with a 30 nm VO2 layer for the remainder of the manuscript, which allows for both high Q-factors and strong resonance modulation. In Fig. 2b, the absolute reflectance modulation across various asymmetries is depicted. It underlines the spectral cleanliness of the switching behaviour since no modes other than the BIC are present, and thanks to the thin $VO_2$ layer, we can switch the metasurface with no noticeable spectral detuning and no off-resonance modulation (Fig. 2b).

We expand our numerical analysis beyond the search for maximum switching in order to also study the intermediate switching states. Fig. 2c captures the progressive degradation of the resonance with increasing $VF$ for a fixed asymmetry $\alpha = 20°$. It reveals a gradual decrease in both Q-factor and amplitude as $VF$ increases, until the resonance almost vanishes. To study the gradual quenching of the BIC in more detail, Fig. 2d presents simulated maps of the electric field enhancement (FE) within the $VO_2$ layer for varying $VF$. For $VF = 0$ (cold phase), there are pronounced FE hotspots at the tips of the ellipses ($FE = 1019$), which reveal the electric dipole in each of the ellipses that form the BIC. While $VF$ increases, the field configuration remains consistent, but the FE eventually drops by almost an order of magnitude at $VF = 1$ (hot phase, $FE = 126$). Fig. 2e depicts the maximal FE extracted from the simulations across all $VF$ in increments of 0.1, demonstrating an almost linear decline. To offer further insight into the attenuation of the mode, we present simulated loss density maps (power absorbed per unit volume[48]) for the varying $VF$ in Fig. 2f. At $VF = 0$, losses manifest in regions with high FE (the tips of the ellipses, consistent with Fig. 2d), a typical behaviour of resonant structures. In contrast, for $VF = 1$, the losses intensify and disperse, peaking in the centers of the ellipses, which indicates non-resonant absorption. The intermediate state with $VF = 0.5$ is a combination of cold and hot phase characteristics with losses both in the center and at the tips. As shown in Fig. 2g, the average loss density within the $VO_2$ layer increases most rapidly for $VF$ between 0.2 and 0.5. A higher $VF$ lead to a saturated loss level, until it declines for $VF = 0.9 - 1$, possibly due to the further decreasing FE.

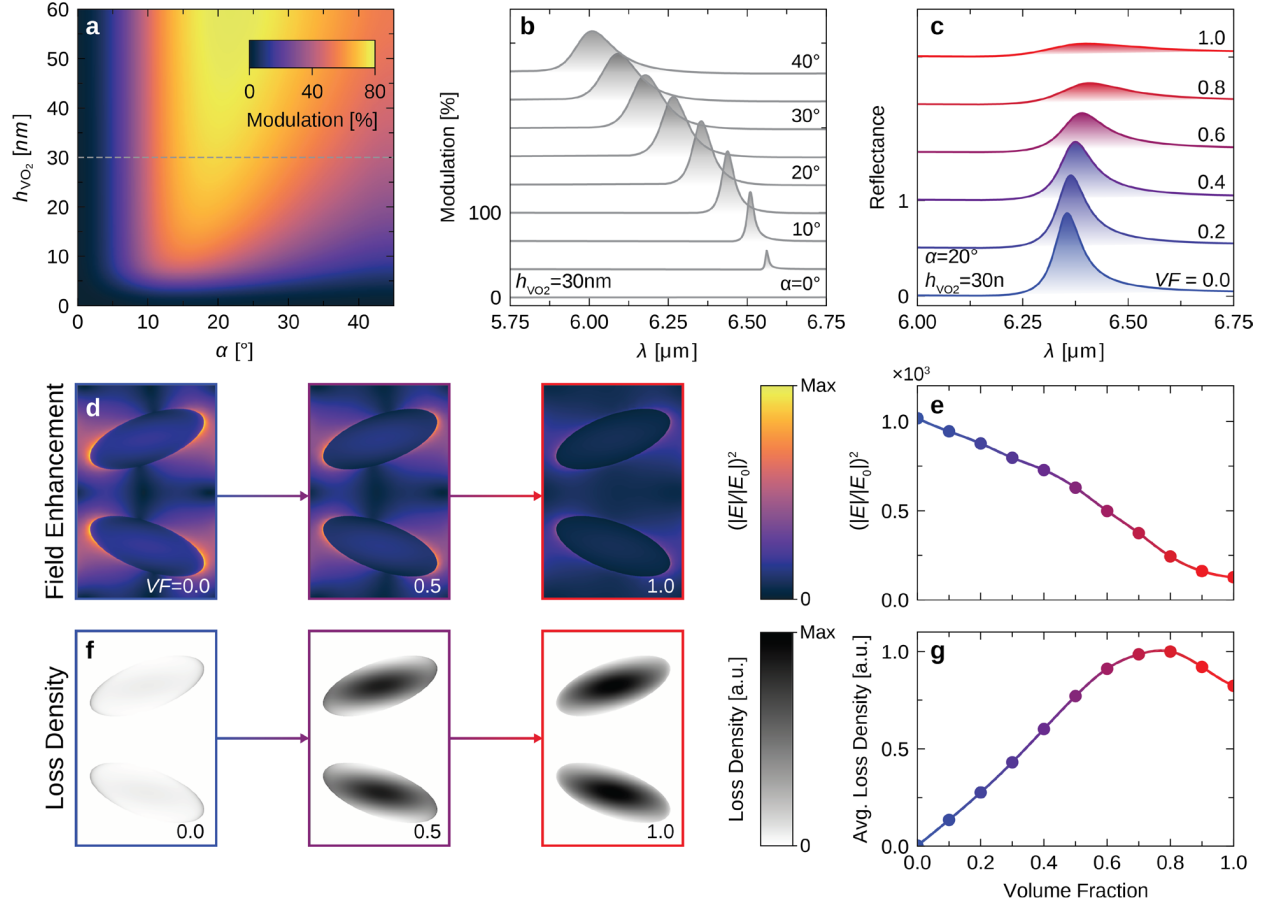

**Figure 2: Numerical studies on the switching behaviour. (a)** Absolute reflectance modulation $\max(R_{\text{cold}} - R_{\text{hot}})$ around 6-6.5 µm as a function of the ellipse tilting angle $\alpha$ and the VO$_2$ layer height $h_{\text{VO}_2}$. A dashed gray line indicates $h_{\text{VO}_2} = 30$ nm, presented in **(b)** where the spectral difference between cold and hot phase reflectance is visualized for various asymmetries, with a vertical offset of 0.33 between the spectra for clarity. **(c)** Spectra of different switching states (different $VF$) for $h_{\text{VO}_2} = 30$ nm and $\alpha = 20°$. Spectra are vertically shifted by 0.5 for clarity. **(d)** Field enhancements in the VO$_2$ layer in the cold ($VF = 0$), intermediate ($VF = 0.5$), and hot ($VF = 1$) phase for the same geometric parameters as in (c). **(e)** Maximal field enhancement for different $VF$. **(f)** Loss density mapping analog to (d). **(g)** Average loss density for different $VF$.

Having verified through simulations that a thin VO$_2$ layer within a BIC resonator can dampen the resonance efficiently and in a controlled manner, we next focus on experimental verification of this effect. We fabricated the BIC metasurface as detailed in the Methods section. All optical measurements were conducted using a Spero microscope (Daylight Solutions), which operates in the wavelength range starting from 5.6 µm, as described in the Methods section. The upper wavelength limit was set to 7.5 µm because, at longer wavelengths, the $SiO_2$ phonon band of the substrate and the capping layer dominates the optical response. Aiming to demonstrate the independent tuning of both intrinsic and radiative losses, we fabricated metasurfaces with varying asymmetries, ranging from $\alpha = 0°$ to $\alpha = 40°$ in the increments of 5°. Fig. 3a presents their reflectance spectra measured at 30°C ($R_{\text{cold}}$) and 100°C ($R_{\text{hot}}$), from which we extracted their absolute reflectance modulation characteristics ($R_{\text{cold}} - R_{\text{hot}}$) in Fig. 3b. Notably, the maximum modulation varies with the asymmetry, reaching almost 50% at $\alpha = 20°$. Generally, the switching process does not compromise the clean low-temperature spectra and produces only minimal off-resonance wavelength modulation. The broadening of the mode at high temperatures is relatively small, while the Q-factor of the reflectance modulation is on par with the resonance in its cold phase. The off-resonance modulation observed around 5.8 µm for $\alpha = 40°$ arises from the BIC interacting with the onset of the first diffraction order at 5.7 µm.

To assess the fine-tuning capabilities of our metasurfaces, we varied the temperature from 25°C to 110°C in 5°C increments. The absolute reflectance modulation amplitude $\max(R_{\text{cold}} - R_{\text{hot}})$ across different asymmetries is plotted in Fig. 3c. The phase transition occurs between 45°C and 75°C, with the most significant changes around 60°C. This offers a sizable 30°C temperature window to fine tune the optical response. In line with the spectra in Fig. 3a,b, the highest maximal modulation is achieved with $\alpha = 20°$, while 10° and 30° show similar results. An asymmetry of 40° yields much lower modulation, and 0° shows negligible change. Driving the metasurface through the whole modulation cycle by heating and cooling reveals a typical hysteresis, as expected for $VO_2$-modulated systems[29]. Our dual-tuning approach lets users prioritize either the *absolute* reflectance modulation discussed so far or the *relative* reflectance modulation $\max\left(\frac{R_{\text{cold}} - R_{\text{hot}}}{R_{\text{cold}}}\right)$. Based on this preference, the optimal structural asymmetry can be chosen, with the peaks for relative and absolute reflectance modulation in our presented structure occurring at $\alpha = 10°$ and $\alpha = 20°$, respectively, see Fig. 3e.

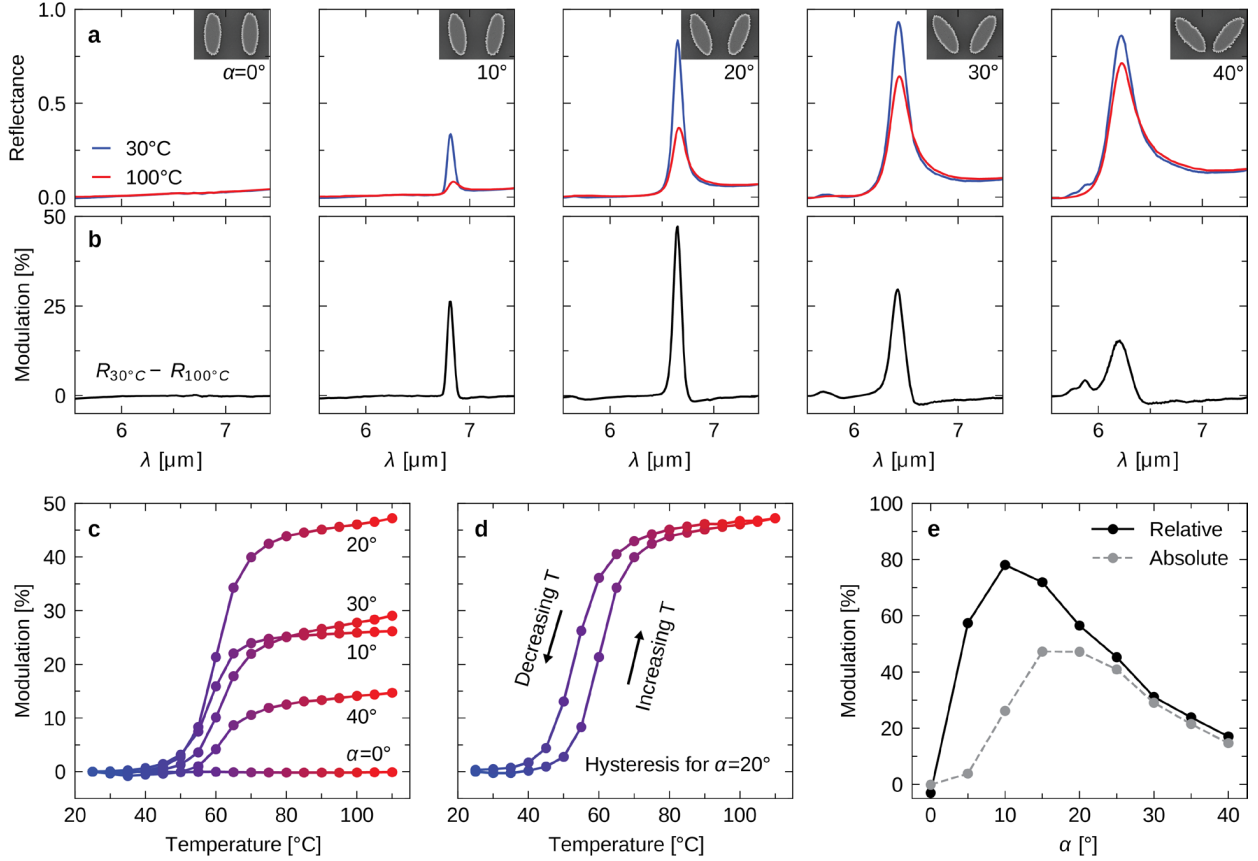

**Figure 3: Experimental switching verification. (a)** Measured reflectance spectra at 30°C (blue) and 100°C (red) for increasing $\alpha$. Insets in the top right corners show SEM images of the respective unit cell with the dimensions of 4400 x 2860 nm². **(b)** Reflectance modulation ($R_{30°C} - R_{100°C}$) extracted from the spectra in (a). **(c)** Temperature-dependent variation of the absolute reflectance modulation $\max(R_{30°C} - R_{100°C})$ for the same asymmetries as in (a) and (b). **(d)** Full hysteresis modulation pattern for $\alpha = 20°$. **(e)** Absolute $\max(R_{30°C} - R_{100°C})$ and relative reflectance modulation $\max\left(\frac{R_{30°C} - R_{100°C}}{R_{30°C}}\right)$ across all asymmetries.

Having demonstrated the experimental switching and fine-tuning capabilities of our metasurfaces, we now take a closer look into the effects the switching has on the resonance characteristics. Our investigation involves spectra from eight distinct asymmetries ($\alpha = 5 - 40°$ in 5° increments) and a temperature series comprising 18 steps ($T = 25 - 110°C$ in 5°C increments). This results in a 2D parameter matrix of 144 unique resonance spectra. To model these resonances, we used temporal coupled mode theory (TCMT), an established framework for interpreting complex light-metasurface interactions[49,50]. Central to TCMT are intrinsic $\gamma_{\text{int}}$ and radiative $\gamma_{\text{rad}}$ loss rates, signifying energy loss within the resonator and energy loss to external channels, respectively. Controlling their ratio is crucial since they dictate the Q-factor, control the FE, and determine the resonance behaviour. In our modelling, we conceptualize our system as a singular resonator coupled to two ports, one symbolizing incoming/reflected light and the other indicating transmitted light, as detailed in Supplementary Note 2. Next, we fit the TCMT model to our experimental data by using unconstrained parameters. Only the intrinsic loss $\gamma_{\text{int}}$ is constrained for all $\alpha$ at a given temperature since it predominantly depends on the intrinsic loss of $VO_2$. In Fig. 4a, we replot the experimental reflectance data from Fig. 3a ($T = 25°C$, varying $\alpha$, solid gray lines) and overlay them with the results of the TCMT fit (dashed black lines).

We extracted the loss rates $\gamma_{\text{rad}}$ (Fig. 4b) and $\gamma_{\text{int}}$ (Fig. 4c) as well as the corresponding total Q-factor (Fig. S3) from all spectra and mapped them across the entire $\alpha$-$T$-parameter space. As $\gamma_{\text{rad}}$ represents the energy radiated out of the resonator, its expected rise with increasing asymmetry is evident, with no dependence on temperature. Conversely, $\gamma_{\text{int}}$, which represents scattering and material losses, exhibits a substantial change with temperature around the phase transition of $VO_2$. Analyzing these loss rates individually underlines a central observation: the two 'tuning knobs' (asymmetry & temperature) modulate the two loss channels ($\gamma_{\text{rad}}$ & $\gamma_{\text{int}}$) independently, allowing selective tailoring of the resonance characteristics. Of special importance is the regime of critical coupling, where $\gamma_{\text{rad}}$ equals $\gamma_{\text{int}}$, marking a sweet spot for a plethora of photonic devices[11,51,52]. It allows optimal energy transfer, heightens sensitivity, and maximizes light absorption. Fig. 4d illustrates the interplay of the two loss rates within our parameter space by plotting their difference $\gamma_{\text{rad}} - \gamma_{\text{int}}$. The brown zone corresponds to undercoupling ($\gamma_{\text{rad}} < \gamma_{\text{int}}$), mainly observed in regions with high $\gamma_{\text{int}}$ (elevated temperatures) and low $\gamma_{\text{rad}}$ (minimal asymmetry). Overcoupling ($\gamma_{\text{rad}} > \gamma_{\text{int}}$), on the other hand, is prevalent at lower temperatures and high asymmetries. The s-shaped dashed black curve marks the domain where resonators are critically coupled.

Our method of integrating a $\gamma_{\text{rad}}$-adjustable BIC with the dynamically tunable $\gamma_{\text{int}}$ of $VO_2$ allows our metasurfaces to modulate between under, over, and critical coupling, dependent on temperature (shown in Fig. 4e for $T = 60°C$) or asymmetry (depicted in Fig. 4f for $\alpha = 10°$). A single metasurface can thus be effortlessly tuned through all coupling regimes, with the steepest changes within the phase transition temperature range of 45°C to 75°C. The path of critical coupling within our 2D parameter space can be further adjusted by simply adjusting the height of the $VO_2$ layer (see Fig. 2a).

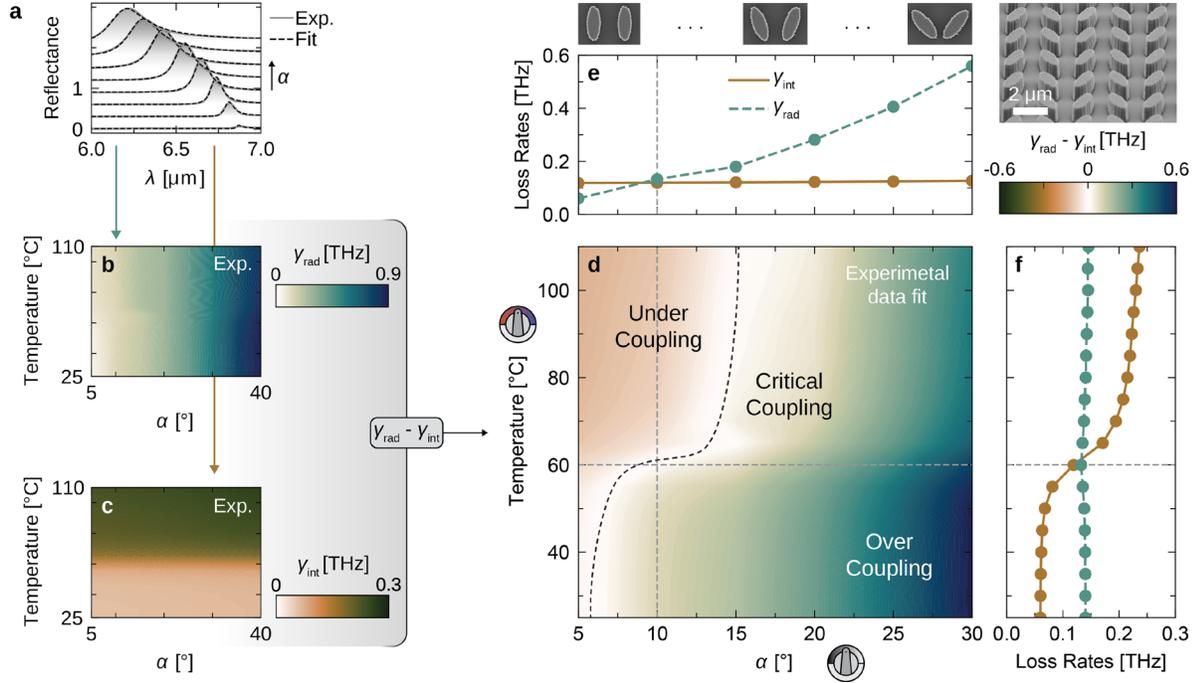

**Figure 4: Insight into experimental data through TCMT analysis:** **(a)** Measured spectra (gray solid line and shading) and TCMT-derived fits (dashed black line) for metasurfaces at 25°C with $\alpha$ values from 5° to 40°. **(b)** Radiative loss $\gamma_{rad}$ and **(c)** intrinsic loss $\gamma_{int}$ extracted from TCMT fits of experimental data across various asymmetries $\alpha$ and temperatures. While $\gamma_{rad}$ exhibits an increase with $\alpha$ and is constant over temperature, $\gamma_{int}$ behaves oppositely: it increases with temperature and is unaffected by $\alpha$. **(d)** Overlay of data from (b) and (c) displaying the difference between the two loss rates $\gamma_{rad} - \gamma_{int}$. Brown shading marks undercoupled resonator regions ($\gamma_{rad} < \gamma_{int}$), green denotes overcoupled regions ($\gamma_{rad} > \gamma_{int}$), and white represents critically coupled resonators, indicated by the s-shaped dashed black curve. **(e)** $\gamma_{rad}$ and $\gamma_{int}$ plotted as functions of the angle $\alpha$ at a temperature of 60°C, based on the data slice highlighted by the dashed horizontal line in (d). **(f)** $\gamma_{rad}$ and $\gamma_{int}$ plotted against temperature for $\alpha = 10°$, corresponding to the data slice marked by the dashed vertical line in (d).

**DISCUSSION**

Our study introduces an innovative method for active control of metasurfaces by harnessing loss-based tuning within a phase-change material in a hybrid BIC metasurface. We utilized the passive Q-factor control of BICs and combined it with the active temperature-sensitive tunability of losses within $VO_2$. This approach enabled us to quench the resonance-based reflectance peak of our metasurfaces, achieving a numerical absolute reflectance modulation of 80% and a relative modulation of 97%. In experiments, these values were realized as 47% and 78%, respectively.

A standout benefit of our approach is that it eliminates the need for thick layers of active materials. Instead, thin layers within the resonator, amounting only to 1/25 of the resonator's volume, are sufficient to suppress the mode. This suppression proves particularly effective at high Q-factors, resulting in a pristine reflectance spectrum free from off-resonant modulations and spectral clutter. Our method provides separate and individual control over the system's key loss channels: the radiative loss, passively controlled by the asymmetry factor of our geometry, and the intrinsic loss, tunable both actively via temperature and passively through the layer thickness of $VO_2$. We further experimentally verified the fine-tuning capability of our metasurfaces, largely attributed to the broad hysteresis of $VO_2$. This property enables robust intermediate switching states, leading to an adjustable modulation of the reflectance. To gain a comprehensive understanding of our system, we mapped the parameter space spanned by the asymmetry parameter $\alpha$ and temperature – the pivotal factors controlling radiative and intrinsic losses. TCMT modelling within this space underline our robust dual control over the loss. Moreover, our findings validate the ability of our structures to transition across all coupling states, from under- to critical-, and then to over-coupling.

This research, while providing substantial insights into loss-based metasurface tuning, also traces a path for applications where precision and controllability of resonances is vital. As the nanophotonics community dives deeper into active metasurfaces, the results presented in this study will likely serve as foundation for more application-based developments. Our dual-control mechanism allows new approaches for advanced sensing of biological samples or gases. Further possible directions include optical switches, fine-tunable filters, thermal radiators, and programmable metasurfaces.

**MATERIALS AND METHODS**

Simulations for our study were performed using the commercial finite element solver CST Studio Suite (Simulia). The software was configured with adaptive mesh refinement, periodic boundary conditions, and operated in the frequency domain.

For sample fabrication, first, 600 nm of Si were deposited onto a fused silica substrate by an ion beam sputtering system with Kaufmann ion sources (Kaufman & Robinson, Inc.) at room temperature by using argon ions at 600 eV. Then, 30 nm of $VO_2$ were deposited by 250 cycles atomic layer deposition (Cambridge NanoTech Fiji 200) running at 150°C. Tetrakis(dimethylamino)vanadium (TDMAV) pre-heated to 87°C and water were used as precursors, that were let into the chamber within 0.6 s pulses with the consecutive waiting times of 8 s and 3 s, respectively. Immediately after the deposition, the sample was annealed in a vacuum tube furnace at 500°C for 5 minutes under 15 sccm flow of oxygen. Finally, 120 nm of $SiO_2$ for the etch mask were deposited onto the sample from a stoichiometric powder in an electron beam evaporator (Bestec, 8 kV, 1 Å/s) at room temperature.

Nanostructuring of the Si – $VO_2$ – $SiO_2$ multilayer film starts with a deposition of a 400 nm layer of positive electron beam resist, ZEP520A (Zeon Corporation), followed by a coating of a conductive polymer, using ESPACER (Showa Denko K.K). Electron beam lithography was carried out at 20 kV using an eLINE Plus system (Raith). Development was achieved with a subsequent bath in amyl acetate and MIBK:IPA in a 1:9 ratio. The polymer mask was used to dry-etch the $SiO_2$ layer and then removed by the Microposit Remover 1165 (Microresist). A second, selective dry-etching process based on $SF_6$ and Argon was applied to etch the $VO_2$ and Si layers. The ellipse's long axis $A$ measures 2460 nm, the short axis $B$ is 980 nm, the x-direction pitch $P_x$ is 4400 nm, and in the y-direction pitch $P_y$ is 2860 nm.

Optical measurements were conducted using a spectral imaging MIR microscope, Spero (Daylight Solutions). The microscope featured a 4× magnification objective ($NA = 0.15$) and provided a 2 mm² field of view. This system features three tunable quantum cascade lasers that consistently cover the 5.6-10.5 μm wavelength range, offering a spectral resolution of 2 cm$^{-1}$. The lasers emit linearly polarized light. Since the SiO$_2$ substrate restricts transmission, all measurements were carried out in reflection mode. To regulate the sample's temperature during experiments, a home-made sample holder was used. This holder consists of a thermally insulated copper contact surface, a temperature sensor, and four heating resistors. Temperature adjustments were controlled by a feedback loop to ensure stability.

**ACKNOWLEDGMENTS**

This work was funded by the Deutsche Forschungsgemeinschaft (DFG, German Research Foundation) under grant numbers EXC 2089/1 – 390776260 (Germany's Excellence Strategy) and TI 1063/1 (Emmy Noether Program), the Bavarian program Solar Energies Go Hybrid (SolTech), and the Center for NanoScience (CeNS), Faculty of Physics/LMU Munich. BUT team was supported by the Grant Agency of the Czech Republic (21-29468S) and the sample fabrication was done in CzechNanoLab Research Infrastructure supported by MEYS CR (LM2023051). S.A.M. acknowledges the Lee-Lucas Chair in Physics and the EPSRC (EP/W017075/1).

**CONFLICT OF INTEREST**

The authors declare no competing financial interest.

# Supplementary Materials for
# Active Loss Engineering in Vanadium Dioxide Based BIC Metasurfaces

**SUPPLEMNTARY NOTES**

**Supplementary Note 1: Effective Medium Approximation**

To model the inhomogeneous nature of the intermediate state of $VO_2$ in our metasurfaces, we utilize the effective medium approximation which is widely adopted in literature[1,2,3]. It calculates the average permittivity of the composite material consisting of domains of $VO_2$ in its hot and cold phases and treats it as a homogeneous medium. This approximation is justified due to the highly subwavelength size of the domains in different phases within the polycrystalline film. The effective permittivity $\varepsilon_{\text{eff}}$ of the $VO_2$ layer can be expressed as a weighted sum of the permittivities of its individual constituents, given by

$$\varepsilon_{\text{eff}} = VF \cdot \varepsilon_{VO_2,\text{hot}} + (1 - VF) \cdot \varepsilon_{VO_2,\text{cold}}$$

As the $VO_2$ undergoes the phase transition, the volume fraction $VF$, defined as $VF = \frac{V_{\text{hot}}}{V_{\text{total}}}$ varies, leading to changes in the effective permittivity $\varepsilon_{\text{eff}}$, see Fig. S4. By integrating this effective medium approximation-based permittivity into our models, we could simulate the metasurfaces behaviour in intermediate states more accurately.

**Supplementary Note 2: TCMT Model**

TCMT provides a straightforward method of modelling resonant structures. Fan et al. presents a general description[4], which we adapt for our single resonance system, coupled to two ports corresponding to reflectance and transmittance. The temporal dynamics of the resonance amplitude $a$ can be written as

$$\frac{da}{dt} = \left(-\frac{\gamma_{rad} + \gamma_{int}}{2} + i\omega_0\right) a + \sqrt{\gamma_{rad}}\, s_{in} \qquad (1)$$

with $\omega_0$ as the resonance frequency of the BIC. $s_{in}$ represents the time dependent input field. Next, we define the relationship between the scattered field $s_{out}$ with the resonance mode and the incoming field by

$$s_{out} = \begin{bmatrix} s_{reflected} \\ s_{transmitted} \end{bmatrix} = S \begin{bmatrix} a \\ s_{in} \end{bmatrix}$$

where $S$ represents the scattering matrix given by

$$S = \begin{bmatrix} -i\sqrt{\gamma_{rad}} & 0 \\ 0 & i\sqrt{\gamma_{rad}} \end{bmatrix}.$$

By using equation (1) under steady state conditions, we are now able to calculate the reflectance $R$ and the transmittance $T$ via their relation to $s_{reflected}$ and $s_{transmitted}$:

$$R = |s_{reflected}|^2 = \gamma_{rad}|a|^2 = \gamma_{rad} \left| \frac{\kappa}{i\omega + \frac{\gamma_{rad} + \gamma_{int}}{2}} \right|^2 |s_{in}|^2$$

$$T = |s_{transmitted}|^2 = \gamma_{rad}|s_{in}|^2$$

$\kappa$ denotes the coupling coefficient between the resonant mode and the external ports. We used this theoretical framework to fit our experimental reflectance spectra to extract parameters like $\gamma_{rad}$ and $\gamma_{int}$. For fitting purposes we normalized $s_{in}$ to 1.

**SUPPLEMNTARY FIGURES**

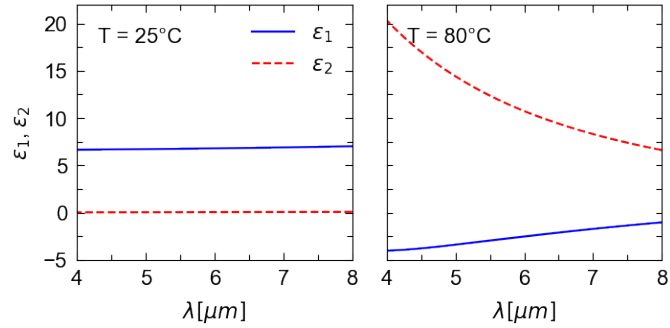

**Figure S1: VO₂ ellipsometry.** Ellipsometry data between 4 and 8 µm for the used 30 nm VO₂ film measured at 25°C for the cold phase and at 80°C for the hot phase. Blue solid (red dashed) lines represent thereal (imaginary) part $\varepsilon_1$ ($\varepsilon_2$) of the VO₂ dielectric function.

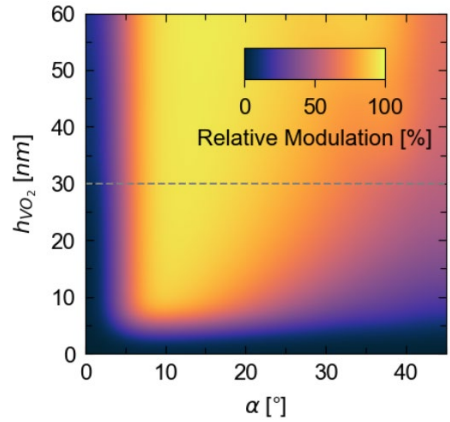

**Figure S2: Numerical Relative Reflectance Modulation.** VO₂ layer height and asymmetry $\alpha$ sweep with the relative reflectance modulation $\max\left(\frac{R_{cold}-R_{hot}}{R_{cold}}\right)$ plotted in color. The maximal value is 97%.

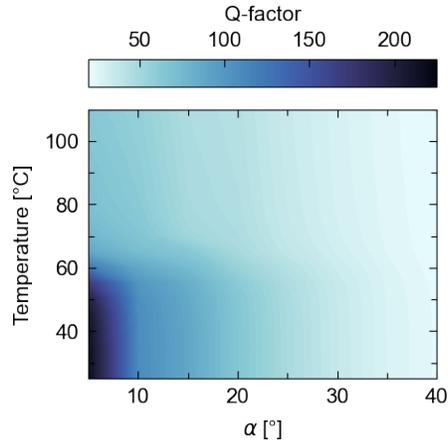

**Figure S3: Q-factor mapping.** Total fitted Q-factor according to our TCMT model, calculated from the two loss rates $\gamma_{\text{rad}}$ and $\gamma_{\text{int}}$ by $Q = \omega_0(\gamma_{\text{rad}} + \gamma_{\text{int}})^{-1}$.

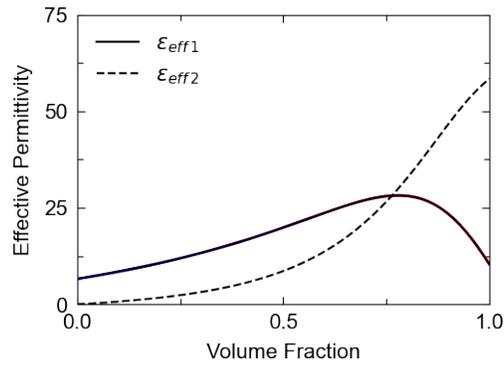

**Figure S4: Effective Permittivity.** Effective permittivity of $VO_2$ for different volume fractions $VF$, using $k$ and $n$ values at a wavelength of 6.5 μm.